%Paper: 9204006
%From: SUMATHI%TIFRVAX.BITNET@tsmi19.sissa.it
%Date: Tue, 14 Apr 92 12:49 IST

\voffset=24pt
%THIS FILE NEEDS TO BE PROCESSED USING THE MACRO PHYZZX.TEX
\overfullrule=0pt
\scrollmode
\input phyzzx

{}~\hfill\vbox{\hbox{TIFR/TH/91-49}\hbox{IP/BBSR/91-38}\hbox{November,
1991}}

\title{CHIRALITY ORDERING OF CHIRAL SPIN LIQUIDS}

\author{D. M. Gaitonde}

\address{Institute of Physics, Sachivalaya Marg, Bhubaneswar 751005,
India}

\author{Dileep P. Jatkar\foot{e-mail address: DILEEP@TIFRVAX.BITNET}
and Sumathi Rao\foot{e-mail address:SUMATHI@TIFRVAX.BITNET}
\foot{Permanent Address:Institute of Physics, Sachivalaya Marg, Bhubaneswar
751005, India.}}

\address{Tata Institute of Fundamental Research, Homi Bhabha Road,
Bombay 400005, India}

\abstract

We study the effect of introducing a weak antiferromagnetic interplanar
exchange coupling in the two dimensional frustrated Heisenberg model.  We
show that a ferromagnetic(FM) ordering of chirality - {\it i.e.}, same
chirality on adjacent planes - is energetically favoured, thus leading to
bulk violation of the discrete symmetries parity($P$) and time
reversal($T$).


\NPrefs
\def\define#1#2\par{\def#1{\Ref#1{#2}\edef#1{\noexpand\refmark{#1}}}}
\def\con#1#2\noc{\let\?=\Ref\let\<=\refmark\let\Ref=\REFS
         \let\refmark=\undefined#1\let\Ref=\REFSCON#2
\let\Ref=\?\let\refmark=\<\refsend}

\define\BEDMUL
J. G. Bednorz and K. A. Muller, Z. Phys. {\bf 64}, 189 (1986); M. K. Wu et
al, Phys. Rev. Lett. {\bf 58}, 908 (1987).

\define\MARWIL
J. March- Russell and F. Wilczek, Phys. Rev. Lett. {\bf 61}, 2066 (1988).

\define\WENZEE
X. G. Wen and A. Zee, Phys. Rev. Lett. {\bf 62}, 2873 (1989).

\define\KALLAUGH
V. Kalmeyer and R. B. Laughlin, Phys. Rev. Lett. {\bf 59}, 2095 (1987).

\define\AFFMAR
I. Affleck and J. B. Marston, Phys. Rev. {\bf B37}, 3744 (1988); J. B.
Marston and I. Affleck, Phys. Rev. {\bf B39}, 11538 (1989).

\define\KOTLIAR
G. Kotliar, Phys. Rev. {\bf B37}, 3664 (1988).

\define\WWZ
X. G. Wen, F. Wilczek and A. Zee, Phys. Rev. {\bf B39}, 11413 (1988).

\define\LZOU
R. B. Laughlin and Z. Zou, Phys. Rev. {\bf B41}, 664 (1990); Z. Zou and R.
B. Laughlin, Phys. Rev. {\bf B42}, 4073 (1990).

\define\SUGEXP
X. G. Wen and A. Zee, Phys. Rev. Lett. {\bf 62}, 2873 (1989); B. I.
Halperin, J. March-Russell and F. Wilczek, Phys. Rev. {\bf 40}, 8726
(1990).

\define\EXPT
R. Kiefl et al, Phys. Rev. Lett. {\bf 64}, 2082 (1990); K. Lyons et al,
Phys. Rev. Lett. {\bf 64}, 2949 (1990); S. Spielman et al, Phys. Rev.
Lett. {\bf 65}, 123 (1990); H. J. Weber et al, Solid State Comm. {\bf 76},
511 (1990).

\define\THRIDFLX
S. B. Libby, Z. Zou and R. B. Laughlin, Nucl. Phys. {\bf B348}, 693
(1991); A. Zee, Mod. Phys. Lett. {\bf B5}, 1331 (1991).

\define\NEUSCAT
S. Chakravarty, in "High Temperature Superconductivity", Los Alamos
Symposium, edited by K. S. Bedell et al, Addison-Wesley (1990).

\define\ROJCAN
A. Rojo and G. S. Canright, Phys. Rev. Lett. {\bf 66}, 949 (1991).

\define\DOPtJ
M. Inui, S. Doniach and M. Gabay, Phys. Rev. {\bf B38}, 6631 (1988); G.
Baskaran and R. Shankar (unpublished).

\define\ANYSUP
A. L. Fetter, G. B. Hanna and R. B. Laughlin, Phys. Rev.  {\bf B39}, 9679
(1989); Y. H. Chen et al, Int. J. Mod. Phys.  {\bf B3}, 1001 (1989).

\define\RICENORI
P. W. Anderson et al, Phys. Rev. {\bf B40}, 8939 (1989); Y. Hasegawa et
al., Phys. Rev. Lett. {\bf 63},907 (1989); P. Lederer et al, Phys. Rev.
Lett.  {\bf 633}, 1519 (1989); F. Nori et al, Phys. Rev. {\bf B41}, 7277
(1990).

\def\bk{\bf k}
\def\bq{\bf q}

\endpage

Ever since the discovery of high $T_c$ superconductors\BEDMUL\ and the
observation of their layered nature, it has been
conjectured\con\MARWIL\WENZEE\KALLAUGH\noc that these materials are
described by a ground state that explicitly violates the discrete
symmetries parity($P$) and time-reversal($T$) macroscopically.
Theoretical interest in these ideas began with the work of Kalmeyer and
Laughlin\KALLAUGH , who (approximately) mapped the Heisenberg model on a
triangular lattice, to a bosonic FQHE problem at filling fraction $\nu =
1/2$, with semionic excitations. An apparently very different line of
investigation was initiated by Affleck-Marston\AFFMAR\ and Kotliar\KOTLIAR
, who introduced the notion of `flux phases'. Working with a frustrated
Heisenberg model on a square lattice, Wen, Wilczek and Zee\WWZ\
generalised the `half-flux' phase of Affleck-Marston to a `quarter-flux'
phase which they called the chiral spin liquid ($CSL$).  They found that
this state explicitly violated $P$ and $T$ macroscopically and in the low
energy long wavelength limit, its effective action led to semionic
statistics, thus corroborating the Kalmeyer-Laughlin picture. More
recently, Laughlin and Zou\LZOU\ have shown that the Gutzwiller projected
$CSL$ state is identical to the Kalmeyer-Laughlin state, paving the way to
a three dimensional generalisation of the physics underlying the FQHE.
The concept of flux phases has also been extended to the doped situation
and generalised flux phases have been shown to be plausible ground states
of the doped $t$-$J$ model\RICENORI.  On the experimental front, many
novel experiments were both suggested\SUGEXP\ and performed\EXPT\ to look
for $P$ and $T$ violation, which appeared to be a robust prediction of all
anyonic theories. But the experimental situation remains confused in the
face of conflicting evidence.

Much of the earlier theoretical work was confined to studies of single
planes. But lately, there have been several attempts to extend flux phase
ideas to the fully three dimensional situation\THRIDFLX.  However, it is
also of both theoretical and experimental importance to incorporate weak
three dimensionality -{\it i.e.} , to study the effect of weak interlayer
couplings -in planar phenomena. We focus on this particular aspect in this
letter. We study the effect of a weak antiferromagnetic interlayer
spin-spin coupling (well motivated by neutron scattering studies\NEUSCAT)
on two dimensional $CSL$ ground states. By perturbatively computing the
correction to the ground state energy, we show that a FM ordering of
chirality on adjacent planes is preferred.  Our work is close in spirit,
but somewhat complementary to the work of Rojo and Canright\ROJCAN, who
studied the ordering of anyons on adjacent planes when a static scalar
potential is introduced between them.  But our work studies the ordering
of the ground states of a microscopic model -the frustrated Heisenberg
model- whereas the starting point of their work involves a gas of anyons.
The two calculations, therefore, cannot be directly compared, since it is
not yet possible to explicitly derive a gas of anyons from the frustrated
Heisenberg model or any other microscopic model.

Let us consider $2p$ planes, each of which has a spin ${\bf S}= 1/2$
sitting on the sites of a square lattice, with Heisenberg
antiferromagnetic interactions between all nearest neighbours within each
plane. The ground state of this model is well known to be Neel ordered.
However, one of the important effects of doping this model with mobile
holes is to induce frustrating interactions\DOPtJ , which cause an
instability towards generalised flux phases. Qualitative features of these
phases are captured by the $CSL$ states of the frustrated Heisenberg
($J$-$J'$) model given by $$ H_{0} = J\sum_{a=1}^{2p}\sum_{<i,j>\in n.n}
{\bf S}_{i}^{a}\cdot{\bf S}_{j}^{a}+ J'\sum_{a=1}^{2p}\sum_{<i,j>\in
n.n.n} {\bf S}_{i}^{a}\cdot{\bf S}_{j}^{a}\eqn\eone $$ which is also an
interesting model in its own right. We shall use this model as our
starting point.  In Eq.\eone, $i$ is the two dimensional site index common
to all the planes and the index $a$ identifies each plane.  As argued in
Ref.\WWZ, the $CSL$ state, characterised by the order parameter $<{\bf
S}_i \cdot {\bf S}_j \times {\bf S}_k>$, where $i,j$ and $k$ are the
vertices of an elementary triangle, is a local minimum of this model for
sufficiently large $J'$. In fact for slightly modified Hamiltonians, it is
a plausible ground state. It is this ground state which has anyonic
excitations and motivates the study of a gas of anyons which forms the
basis of theories of anyon superconductivity\ANYSUP.  In this letter, we
shall focus our attention on the mean field description of this $CSL$
state.

We use a fermionic description for the spins given by $$ {\bf S}_i^{c,a} =
\sum_{\alpha,\beta} c_{i\alpha}^{a \dag} {\bf \sigma}_{\alpha\beta}
c_{i\beta}^a \quad{\rm and}\quad {\bf S}_{i}^{d,a} = \sum_{\alpha, \beta}
d_{i\alpha}^{a \dag} {\bf \sigma}_{\alpha\beta} d_{i\beta}^a,\eqn\etwo $$
where we have distinguished alternate planes by the nomenclature of the
fermions as $c$ and $d$ planes. Also, for clarity and brevity of notation,
we shall henceforth drop the index $a$ and the summation over $a$ which
now goes from $1$ to $p$, since odd and even planes have been
distinguished.  In terms of the fermions, the Hamiltonian $H_0$ can be
rewritten, after a Fierz transformation followed by a Hubbard-Stratanovich
transformation as $$
\eqalign{H_0 &= \sum_{\{i,j\},\alpha} [\chi_{ij}^{c} c_{i\alpha}^{\dag}
c_{j\alpha} + \chi_{ij}^{d} d_{i\alpha}^{\dag} d_{j\alpha} + h.c.] +
\sum_{i,\alpha} [a_{0i}^{c} (c_{i\alpha}^{\dag} c_{i\alpha} - 1) +
a_{0i}^{d} (d_{i\alpha}^{\dag} d_{i\alpha} - 1)]\cr
&~~~~~~~~~+ {2\over J} \sum_{<i,j>\in n.n} \chi_{ij}^{c,d\dag} \chi_{ij}^{c,d}
+ {2\over J'} \sum_{<i,j>\in n.n.n} \chi_{ij}^{c,d\dag} \chi_{ij}^{c,d}}
\eqn\ethree
$$
where $\chi_{ij}^{c,d}$ are the Hubbard-Stratanovich fields. The notation
$\{i,j\}$ in the first summation stands for summation over both nearest
and next nearest neighbours.

Following WWZ\WWZ, we introduce the mean field ansatz for the chiral spin
liquid state for both the $c$ and $d$ planes. For the nearest neighbour links,
$$
\langle\chi_{i,i\pm{\hat x}}^{c,d}\rangle = g e^{i\pi/4} \quad {\rm and}\quad
\langle\chi_{i,i\pm{\hat y}}^{c,d}\rangle = g e^{-i\pi/4} \eqn\efoura
$$
where $i$ here, is a site on the odd sublattice. However, for the diagonal
links, the WWZ ansatz admits a two-fold degeneracy corresponding to the
two possible chiralities - {\it i.e.}, the flux through each elementary
plaquette, which is now a triangle, could either be positive or negative.
Since we wish to study the ordering of chiralities on different planes, we
allow for independent chiralities on the two planes. Thus, the n.n.n links
are described by
$$
\langle\chi_{i,i-{\hat x}+{\hat y}}\rangle = \delta_{i} f^{c,d}\quad {\rm and}
\quad\langle\chi_{i,i-{\hat x}-{\hat y}}\rangle = -\delta_{i}f^{c,d}\eqn\efourb
$$
where $\delta_i = +(-)1$ for $i$ belonging to the even (odd) sublattice.
$f^c = f^d$ implies that the chiralities on the adjacent $c$ and $d$
planes are the same (FM ordering) and $f^c = -f^d$ implies an AFM ordering.
In the absence of any interplanar coupling, the two possibilities
obviously remain degenerate. Finally, for the Lagrange multiplier fields
we have
$$
\langle a_{0i}^{c}\rangle = \langle a_{0i}^{d}\rangle = 0, \eqn\efourc
$$
so that the fermions are no longer subject to the `no double occupancy'
constraint at each site. The constraint is now enforced only on the average.

Notice that the mean field ansatz for the $CSL$ state divides
the square lattice on each plane into two sublattices. Thus, we may take
the spatial Fourier transformations separately for the odd and even
sublattices, with respect to a 2-d wave vector ${\bk}$ which now runs over
the reduced Brillouin zone ($RBZ$). Hence, the momentum space mean field
Hamiltonian is given by
$$
H^{MF} = \sum_{{\bk}\in RBZ,\alpha} [\psi_{{\bk}\alpha}^{c\dag} h_{{\bk}}^c
\psi_{{\bk}\alpha}^c + \psi_{{\bk}\alpha}^{d\dag} h_{{\bk}}^d
\psi_{{\bk}\alpha}^d] \eqn\efive
$$
where $\psi_{{\bk}\alpha}^c = (c_{{\bk}\alpha}^o, c_{{\bk}\alpha}^e)$ and
$\psi_{{\bk}\alpha}^d = (d_{{\bk}\alpha}^o, d_{{\bk}\alpha}^e)$, $o$ and $e$
stand for odd and even respectively and
$$
\eqalign{h_{{\bk}}^{c,d} &\equiv \pmatrix{\epsilon_{\bk}^{c,d}&\Delta_{\bk}\cr
\Delta_{\bk}^{*}&-\epsilon_{\bk}^{c,d}\cr}\cr
&= \pmatrix{2f^{c,d} [\cos(k_{x}+k_{y})-\cos(k_{x}-k_{y})]&
2g[-i \cos(k_{x})+\cos(k_{y})]\cr
2g[i \cos(k_{x})+\cos(k_{y})]&-2f^{c,d}[\cos(k_{x}+k_{y})-\cos(k_{x}-k_{y})]
\cr}}
\eqn\esix
$$
This Hamiltonian can be diagonalised by a unitary transformation yielding
$$
h_{{\bk},{\rm diag}}^{c,d} = \pmatrix{-E_{\bk}^{c,d}&0\cr
0&E_{\bk}^{c,d}} \eqn\eseven
$$
with $E_{\bk}^{c,d}=(|\Delta_{\bk}|^{2}+
(\epsilon_{\bk}^{c,d})^{2})^{1/2}$
in terms of the transformed variables $(\gamma_{{\bk}\alpha}^V,
\gamma_{{\bk}\alpha}^C)$ and $(\eta_{{\bk}\alpha}^V,
\eta_{{\bk}\alpha}^C)$  for the valence band(V) and conduction band(C) fermions
in the $c$ and $d$ planes respectively. The ground state has the valence
band completely filled in both the planes and its energy, in terms of the
mean field variables, is given by
$$
E_{0}^{MF} = {2\over J} \sum_{i\in odd} \sum_{j(i)\in n.n} g^2 + {2\over J'}
\sum_{i\in odd} \sum_{j(i)\in n.n.n} f^2 - 2 \sum_{{\bk}\in RBZ} E_{{\bk}} .
\eqn\eeight
$$

In the absence of any interplanar coupling, the FM and AFM orderings of
chirality remain degenerate. To lift the degeneracy, we introduce a weak
interlayer Heisenberg antiferromagnetic coupling given by
$$
H_{int} = J''\sum_{i} {\bf S}_i^c \cdot {\bf S}_i^d \eqn\enine
$$
where ${\bf S}_i^c$ and ${\bf S}_i^d$ refer to the spins on the $c$ and
$d$ planes respectively. Such an interaction is particularly appropriate
for the copper oxide systems and leads to 3-d Neel ordering in
the undoped insulating phase.  $J''$ has been estimated from neutron
scattering experiments\NEUSCAT , to be about five orders of magnitude less
than the in-plane coupling $J$. We treat $H_{int}$ as a static perturbing
potential between the two species of fermions on adjacent planes. This is
accomplished by taking momentum space Fourier transformations with respect
to a 2-d wave vector. Thus, despite the extension of the problem into the
third dimension, inter-layer particle transfers are avoided and the
essential layered nature of the original problem is retained. The relative
weakness of $J''$ with respect to $J$ justifies this approach.

In the fermionic representation,
$$
H_{int} = {J''\over  2}\sum_{i,\alpha,\beta} c_{i\alpha}^{\dag} c_{i\beta}
d_{i\beta}^{\dag} d_{i\alpha}\eqn\eten
$$
which, when Fourier transformed with respect to 2-d wave vectors,
becomes
$$
H_{int} = {J''\over N}\sum_{{\bk},{\bk}',{\bq}\in RBZ}
[c_{{\bk}+{\bq}\alpha}^{o\dag}c_{{\bk}\beta}^{o}d_{{\bk}'
-{\bq}\beta}^{o\dag}d_{{\bk}'\alpha}^{o}
+ c_{{\bk}+{\bq}\alpha}^{e\dag}c_{{\bk}\beta}^{e} d_{{\bk}'
-{\bq}\beta}^{e\dag} d_{{\bk}'\alpha}^{e}]. \eqn\eeleven
$$
Notice that a change in momentum in the $c$ plane is compensated by an
opposite change in momentum in the $d$ plane. We now evaluate the total
ground state energy, treating $H_{int}$ as a perturbation, for the two
cases of FM and AFM orderings of chirality. The unperturbed ground state
is given by
$$
|{\rm Ground\quad
state}\rangle=\prod_{{\bk},\alpha}\gamma_{{\bk}\alpha}^{V\dagger}
|0\rangle
\otimes\prod_{{\bk},\alpha}\eta^{V\dagger}_{{\bk}\alpha}|0\rangle
\eqn\etwelve
$$
and the unperturbed ground state energy is given in Eq.\eeight\ . The FM
ground state has $f^{c} = f = f^{d}$, whereas the AFM ground state has
$f^{c} = f = -f^{d}$, so that the Boguliobov transformation coefficients
and hence the definition of the transformed fermions $\eta_{{\bk}\alpha}^{V,C}$
differ in the two cases. It is now
straightforward to rewrite $H_{int}$ in terms of the transformed fermions
and compute $E_{FM} - E_{AFM}$.

 At first order, we find that
$$
E^{(1)}_{FM} - E^{(1)}_{AFM} = {2J''\over
N}(\sum_{k}{\epsilon_{{\bk}}\over E_{{\bk}}})^{2} = 0,\eqn\ethirteen
$$
since $\epsilon_{{\bk}}/E_{{\bk}}$ is odd under reflection about the
$k_{y}$-axis and the summation over ${\bk}$ includes both positive and
negative $k_{x}$. This result is easily understood, since at first order,
the only term in $H_{int}$ that contributes involves no momentum transfer
${\bq}$ hence, the two planes are essentially independent and the
degeneracy between FM and AFM orderings of chirality is not lifted. At
second order too the degeneracy is not lifted. We find that
$$
E^{(2)}_{FM} - E^{(2)}_{AFM} = ({J''\over 2N})^{2}\sum_{{\bk},{\bk}',{\bq}}
{1\over {E_{{\bk}}+E_{{\bk}'}+E_{{\bk}+{\bq}}+E_{{\bk}'-{\bq}}}}
({\epsilon_{{\bk}'-{\bq}}\over
E_{{\bk}'-{\bq}}}-{\epsilon_{{\bk}'}\over E_{{\bk}'}})
({\epsilon_{{\bk}+{\bq}}\over
E_{{\bk}+{\bq}}}-{\epsilon_{{\bk}}\over E_{{\bk}}}).\eqn\efourteen
$$
Making the changes ${\bk} \rightarrow {\bk} - {\bq}$, ${\bq}\rightarrow
-{\bq}$, and ${\bk}'\rightarrow -{\bk}'$ succesively in the dummy
variables, and using $\epsilon_{-{\bk}} = \epsilon_{{\bk}}$,
$\Delta_{-{\bk}} = \Delta_{{\bk}}$ and $E_{-{\bk}} = E_{{\bk}}$, we find
that $$ E^{(2)}_{FM}-E^{(2)}_{AFM} = -(E^{(2)}_{FM}-E^{(2)}_{AFM}) =
0.\eqn\efifteen $$ However, the third order contribution does lift the
degeneracy and is given by $$ E^{(3)}_{FM} - E^{(3)}_{AFM} = E_{A} + E_{B}
+ E_{C}\eqn\esixteen $$ where $$
\eqalign{E_{A} &= -{J''^{3}\over N^{3}}\sum_{{\bk},{\bk}',{\bq},{\bq}'}
({e_{{\bk}+{\bq}}
+ e_{{\bk}+{\bq}+{\bq}'} - e_{{\bk}} -
e_{{\bk}}e_{{\bk}+{\bq}}e_{{\bk}+{\bq}+{\bq}'}\over {E_{\bk}+E_{{\bk}+{\bq}}+
E_{{\bk}'}+E_{{\bk}'-{\bq}}}}) \cr &~~~~~[{e_{{\bk}'}+e_{{\bk}'+{\bq}'}-
e_{{\bk}'-{\bq}}
-e_{{\bk}'} e_{{\bk}'+{\bq}} e_{{\bk}'-{\bq}} \over {E_{{\bk}}+E_{{\bk}+{\bq}
+{\bq}'}+E_{{\bk}'+{\bq}'}+E_{{\bk}'-{\bq}} }}\cr
&~~~~-{e_{{\bk}'-{\bq}}+e_{{\bk}'-{\bq}-{\bq}'}-e_{{\bk}'}-e_{{\bk}'}
e_{{\bk}'-{\bq}-{\bq}'} e_{{\bk}'-{\bq}} \over {4(E_{{\bk}}+
E_{{\bk}+{\bq}+{\bq}'}+E_{{\bk}'}+E_{{\bk}'-{\bq}-{\bq}'}) }}],}
\eqn\eseventeen
$$
$$
\eqalign{E_{B} &= -{J''^{3} \over 2 N^{3}} \sum_{{\bk},{\bk}',{\bq},{\bq}'}
{1\over (E_{{\bk}}+E_{{\bk}+{\bq}}+E_{{\bk}'}+E_{{\bk}'-{\bq}})
(E_{{\bk}}+E_{{\bk}+{\bq}+{\bq}'}+E_{{\bk}'+{\bq}'}+E_{{\bk}'-{\bq}})}\cr
&~~~~~~~~~~~~~~~~~~~~[e_{{\bk}+{\bq}} e_{{\bk}'} (\delta_{{\bk}}^{*}
\delta_{{\bk}+{\bq}+{\bq}'}
\delta_{{\bk}'+{\bq}'}^{*} \delta_{{\bk}'-{\bq}} + h.c.)\cr
&~~~~~~~~~~~~~~~~~~~+e_{\bk} e_{{\bk}'-{\bq}} (\delta_{{\bk}+{\bq}}^{*}
\delta_{{\bk}+{\bq}+{\bq}'}
\delta_{{\bk}'} \delta_{{\bk}'+{\bq}'}^{*} + h.c.)\cr
&~~~~~~~~~~~~~~~~~~~+e_{{\bk}+{\bq}+{\bq}'} e_{{\bk}'+{\bq}'}
(\delta_{\bk} \delta_{{\bk}+{\bq}}^{*}
\delta_{{\bk}'} \delta_{{\bk}'-{\bq}}^{*} +h.c.)],} \eqn\eeighteen
$$
and
$$
\eqalign{E_{C} &= {J''^{3} \over 8 N^{3}} \sum_{{\bk},{\bk}',{\bq},{\bq}'}
{1\over (E_{\bk}+E_{{\bk}+{\bq}}+E_{{\bk}'}+E_{{\bk}'-{\bq}})
(E_{\bk}+E_{{\bk}+{\bq}+{\bq}'}+E_{{\bk}'}+E_{{\bk}'-{\bq}-{\bq}'})} \cr
&~~~~~~~~~~~~~~~~~~~[e_{{\bk}+{\bq}} e_{{\bk}'-{\bq}} (\delta_{\bk}
\delta_{{\bk}+{\bq}+{\bq}'}^{*}
\delta_{{\bk}'} \delta_{{\bk}'-{\bq}-{\bq}'}^{*} + h.c.)\cr
&~~~~~~~~~~~~~~~~~~+e_{\bk} e_{{\bk}'} (\delta_{{\bk}+{\bq}}
\delta_{{\bk}+{\bq}+{\bq}'}^{*}
\delta_{{\bk}'-{\bq}} \delta_{{\bk}'-{\bq}-{\bq}'}^{*} + h.c.) \cr
&~~~~~~~~~~~~~~~~~~+e_{{\bk}+{\bq}+{\bq}'} e_{{\bk}'-{\bq}-{\bq}'}
(\delta_{\bk}^{*} \delta_{{\bk}+{\bq}}
\delta_{{\bk}'}^{*} \delta_{{\bk}'-{\bq}} + h.c.)].}\eqn\enineteen
$$
(Here, $e_{\bk} = \epsilon_{\bk}/E_{\bk}$ and $\delta_{\bk} =
\Delta_{\bk}/E_{\bk}$ for any momentum ${\bk}$.) The {\bk}- summations in
Eqs.\eseventeen,
\eeighteen\ and \enineteen\
were performed numerically using a Monte
Carlo routine, for different values of $J'/J$, with the corresponding
values of $f/J$ and $g/J$ being obtained by minimising $E_0$ in
Eq.\eeight\ with respect to $f$ and $g$. Our numerical results are
tabulated below.

\advance\hsize by -1.5in
%\advance\hoffset by -.5in
$$
\hskip 0.75in\vbox{\baselineskip=28pt\centerline{Table~I}\vskip12pt\hrule
\vskip12pt{\settabs\+ aaaaaaaaaaa & aaaaaaaaaaaaa & aaaaaaaaaaa &
aaaaaaaaaaaaaaaaaaaaaaa \cr
\+ ~~$J'/J$ & ~~$f/J$ & ~~$g/J$ & ~~~~($E_{FM}^{(3)} - E_{AFM}^{(3)}$)/$J$ \cr}
\vskip12pt\hrule\vskip12pt
{\settabs\+ aaaaaaaaaaa & aaaaaaaaaaaaa & aaaaaaaaaaa  &
aaaaaaaaaaaaaaaaaaaaaaa \cr
\+ ~~0.50 & ~~0.015 & ~~0.23 & ~~$-$ 1.7 $\times 10^{-4} N (J''/ J)^{3}$ \cr
\+ ~~0.55 & ~~0.024 & ~~0.24 & ~~$-$ 2.9 $\times 10^{-4} N (J''/ J)^{3}$ \cr
\+ ~~0.60 & ~~0.035 & ~~0.24 & ~~$-$ 4.6 $\times 10^{-4} N (J''/ J)^{3}$ \cr
\+ ~~0.65 & ~~0.046 & ~~0.23 & ~~$-$ 6.8 $\times 10^{-4} N (J''/ J)^{3}$ \cr
\+ ~~0.70 & ~~0.060 & ~~0.23 & ~~$-$ 8.2 $\times 10^{-4} N (J''/ J)^{3}$ \cr
\vskip12pt\hrule }}
$$
\advance\hsize by 1.5in
%\advance\hoffset by .5in

Thus, for any value of $J'/J$ for which the $CSL$ state is a
local minimum, and for $J'' > 0$, (which is the case for copper
oxides), the FM ordering of chirality is energetically favoured. Notice
that the energy difference is an extensive quantity and scales linearly
with $N$. In fact, using typical values for $La_{2} Cu O_{4}$, ( $J =
1200^\circ K$ and $J'' = 0.03^\circ K$), and assuming $N \sim 10^{15}$,
$E_{AFM} - E_{FM}$ ranges between $3.2^\circ K$ and $15.4^\circ K$ for $J'/J$
between 0.5 and 0.7. Thus, despite the weakness of the interlayer
coupling, its potency is effectively increased by its extensivity. Hence,
at low enough temperatures, the weak Heisenberg antiferromagnetic
interlayer coupling has sufficient strength
to tilt the scales in favour of a FM ordering of chirality.

In our calculation, we have completely ignored gauge field fluctuations
-{\it i.e.,} the phase fluctuations of the order parameter $\chi_{ij}$ and
the fluctuations of the Lagrange multiplier field $a_{0i}$. These
fluctuations could lead to a substantial contribution  to the ground state
energy.  However, they cannot lift the degeneracy between the FM and AFM
orderings of chirality, since they only act within each plane. Thus, as
long as these fluctuations do not destabilise the mean field ground state,
$E_{FM} - E_{AFM}$ and consequently, the ordering of chirality is
determined only by the interplanar coupling.

We have worked within the framework of the $J$-$J'$ model, which is a
limiting case of the
$t$-$t'$-$J$-$J'$ model. However, we
expect the qualitative aspects of our result - {\it i.e.,} the tendency
towards FM ordering of chirality - to be valid even for the generalised
flux phases of the doped $t$-$J$ model,
at least for low doping. Notice that our result suggests
that bulk $P$
and $T$ violation is an inescapable consequence of $CSL$ ground states of
models that are relevant to high $T_c$ superconductors. Moreover, despite
the controversy regarding the observation of local $P$ and $T$ violation,
bulk $P$ and $T$ violation has certainly been ruled out in the cuprate
compounds\EXPT. Hence, our calculation disfavours models with $CSL$ ground
states as candidates to describe the doped Mott insulating phases of the
copper oxides.

\ack
We would like to thank A. Sen for many illuminating conversations at
various stages of this work. We would also like to thank G. Baskaran and
R. M. Godbole for useful suggestions.

\refout
\end